\documentclass[fleqn,usenatbib]{mnras}

\usepackage[T1]{fontenc}

\DeclareRobustCommand{\VAN}[3]{#2}
\let\VANthebibliography\thebibliography
\def\thebibliography{\DeclareRobustCommand{\VAN}[3]{##3}\VANthebibliography}

\usepackage{graphicx}	
\usepackage{amsmath}	
\usepackage{amssymb}	
\usepackage{scalefnt}

\usepackage{newtxtext,newtxmath}

\def\lsim{\;\raise0.3ex\hbox{$<$\kern-0.75em\raise-1.1ex\hbox{$\sim$}}\;}
\def\gsim{\;\raise0.3ex\hbox{$>$\kern-0.75em\raise-1.1ex\hbox{$\sim$}}\;}
\newcommand{\vect}[1]{\boldsymbol{#1}}
\newcommand{\muG}{\,$\mu$G\,}
\newcommand{\kms}{\,km\,s$^{-1}$\,}

\newcommand{\Msun}{\,M$_{\odot}$\,}
\newcommand{\MsunYr}{\,M$_{\odot}$\,yr$^{-1}$\,}
\newcommand{\pccell}{\,pc\,cell$^{-1}$\,}

\newcommand{\textmc}[1]{\textsc{\scalefont{1.15}#1}}
\title[Supernova inside the core of a YMSC]{Core-collapse supernova inside the core of a young massive star cluster: \\ 3D MHD simulations}

\author[D. V. Badmaev, A. M. Bykov, M. E. Kalyashova]{
D. V. Badmaev,\thanks{E-mail: danirbadmaev.astro@gmail.com}
A. M. Bykov, M. E. Kalyashova
\\
Ioffe Institute, 26 Politekhnicheskaya St, 194021 Saint Petersburg, Russia
}

\date{Accepted XXX. Received YYY; in original form ZZZ}

\pubyear{2023}

\begin{document}
\label{firstpage}
\pagerange{\pageref{firstpage}--\pageref{lastpage}}
\maketitle

\begin{abstract}
Young massive stars in compact stellar clusters could end their evolution as core-collapse supernovae a few million years after the cluster was built. The blast wave of a supernova propagates through the inner cluster region with multiple stellar winds of young luminous stars. We present the results of 3D magnetohydrodynamic simulations of the plasma flows produced by a supernova event inside a cluster with a population of massive stars similar to that in Westerlund 1. We followed its evolution over a few thousand years (i.e. a few shock crossing times). The plasma temperature, density and magnetic field,  which are highly disturbed by supernova event, relax to values close to the initial over the studied period. The relaxation time of a cluster is a few thousand years, which is a sizeable fraction of the period between the successive supernova events for a massive cluster of a few million years age. The spectra of the cluster diffuse X-ray emission simulated here should be representative for the galactic and extragalactic young massive clusters. The resultant magnetic fields are highly intermittent, so we derived the volume filling factors for a set of magnetic field ranges. Highly amplified magnetic fields of magnitude well above 100 $\mu$G fill in a few per cent of the cluster volume, but still dominate the magnetic energy. The structure of the magnetic fields and high velocity plasma flows with shocks in the system are favorable for both proton and electron acceleration to energies well above TeV.
\end{abstract}




\section{Introduction}
\label{sec:intro}

Young massive star clusters (YMSCs) host rich and dense populations of luminous massive stars which have short lifespans of a few Myr. Some of these stars end their lives as core-collapse supernovae (SNe) releasing $\sim 10^{51}$\,erg of kinetic energy into the surrounding medium. The interaction of powerful winds from luminous stars can create a complex and highly magnetized medium with numerous shocks. When the forward shock of a supernova remnant (SNR) propagates through the circumstellar medium of massive stars it is expected to further compress the gas and amplify the magnetic field. In this paper, we aimed to investigate the dynamics of a supernova remnant in the context of a YMSC and its impact on the inner cluster environment.

YMSCs play an important role in star-forming regions being the powerful sources of ionising radiation, kinetic energy and momentum, which affect their parent clouds \citep[see for review][]{Kru19,Kra20} and the interstellar medium (ISM) \citep[e.g.][]{Romero20}. The burst mode of episodic accretion can explain the high-mass star formation processes \citep[see e.g.][]{burst19,burst21}. In the galaxies with high star formation rate a substantial fraction of massive stars are born in YMSCs \citep{Ada20}. Moreover, they were suggested recently as likely places for the formation of binary black holes which are the sources of gravitational waves \citep[e.g.][]{binary_BHs23}. Both stellar and diffuse emission from some clusters was thoroughly studied over the whole range of the electromagnetic waves including the high energy photons.

Diffuse X-ray emission from hot plasma produced by colliding stellar winds (SWs) and SNe was revealed in YMSCs \citep[see][]{Mun06,Wd2_Townsley19,Kavanagh20,Sasaki22,Wd2_artXC23}. Some clusters were also detected in very high energy gamma-rays \citep[e.g.][]{Wd2_Ahar07,Wd1_HESS22,Ohm_clusters10,Wd2_Yang18} which is a clear indication of efficient cosmic ray (CR) acceleration processes in these sources which can be studied in detail with the forthcoming Cherenkov Telescope Array observatory \citep[see e.g.][]{CTA23,CTA23a}. CRs accelerated in the clusters can penetrate deep into the cores of nearby molecular clouds providing the gas heating and ionization. This effect is known to be important to both star formation process and interstellar chemistry  \citep[e.g.][]{Draine_ISM11,padovani20}. Indeed, due to the high extinction the UV and soft X-ray photon fluxes drop drastically inside the dense cloud clumps and the low-energy CRs are the main ionization agent. Being localized inside the molecular clouds and rather long-lived (up to some Myr) the low-energy CR sources associated with YMSCs can be an important feedback agent in the star formation process in galaxies. On the other hand, the high-energy protons accelerated in YMSCs interacting with the surrounding matter can contribute to the observed high-energy neutrino fluxes \citep[see e.g.][]{bego15}. An adequate interpretation of the high-energy diffuse emission requires detailed modeling of the energy partition processes in the complex plasma flows, produced by the colliding winds of young massive stars and SN events. Models of particle acceleration in YMSCs at the evolution phase, dominated by the powerful winds of OB and Wolf Rayet (WR) stars and SNe \citep[it is discussed in][]{Byk14,Gupta20,Morlino21,Vieu23,Gabici23}, rely on the dynamics of the shocks of different strengths and scales, as well as on the magnetic fields in clusters and their vicinity. The hydrodynamic modeling of the structure of YMSCs and their impact on the galactic environment was performed extensively \citep[see e.g.][]{Chv85,SH03,Torres07,RP13,TT17,Bla20}.  To study the structure and evolution of magnetic fields in YMSCs at the stage dominated by powerful stellar winds we performed a 3D magnetohydrodynamic (MHD) simulation of the quasi-stationary medium inside a YMSC core \citep[see Paper~I:][]{Bad22}. In Paper I we discussed the formation of cluster-scaled magnetic field filaments, where the amplitude of the field reaches $\gsim 100$\muG. This model did not include possible supernova events and thus it is only applicable for a period of time ($<10$\,kyr) when there are no strong and rapid ($t_{\mathrm{dyn}}\lesssim1$\,kyr) perturbations as the stellar system evolves. 

\citet{Chv82} and \citet{Nad85} proposed a self-similar solution for an SNR freely expanding at the earliest stages of evolution. This solution considers the expansion of the supernova ejecta into the surrounding medium and can be applied to different types (Ia, Ib/c-II) of SNe occurring in either a uniform or wind-blown circumstrellar medium. Hydrodynamic modeling of supernova remnant evolution inside a centrally symmetric wind-blown bubble was performed, considering both stationary \citep{1982SvAL....8..361B,CL89,Tenorio90} and moving massive progenitor \citep{roszyczka93}, demonstrating complex behaviour of the ejected material and velocity fields. The origin of strong magnetic fields in young SNRs has been investigated through 2D MHD simulations by \citet{JN96}. Models with a smooth transition between the free-expansion and Sedov-Taylor phases have also been introduced \citep[see][]{TM99}. The evolution of a SN shock in the bubble of the massive progenitor star which accounted for different mass-loss regimes at the evolution phases of O star, red supergiant, and WR was simulated by \citet[]{Dwarkadas07}. MHD models have also been applied by \citet{Mar18} to the problem of CR acceleration. The effect of magnetic fields in the wind of the progenitor star of SN1987a has been studied in detail by \citet{Orlando_SN1987a19}. Recent developments include MHD models for the dynamics of core-collapsed SNe \citep[][]{Pet21}, connecting the early phases of supernovae to the observed morphology of SNRs in multi-wavelength observations \citep[see e,g.][]{Vink20, ferrand21}. Some of the latest 3D HD/MHD models demonstrate the connection between the dynamical properties of SNRs and the internal structure of their progenitor stars, as well as the circumstellar medium \citep[see][]{Orl20,Orl22}. Non-spherical morphology of SNRs propagating in anisotropic wind-blown bubbles shaped by interstellar magnetic fields has also been investigated by \citet{Meyer_anisobubble22}. Recently, \citet{Mey23} have studied the mixing of materials in magnetized core-collapse SNRs moving through the ISM over a period of $\sim10$ kyr. Additionally, collisions between supernova remnants and stellar winds from nearby massive stars have been simulated using both hydrodynamic and MHD models \citep[e.g.][]{Bad21,Vel03,Vel23}.

Given the number of massive stars ($M>25\,M_{\odot}$) in a YMSC core $\sim 100$ and considering  YMSCs of ages below 5 Myr, one could expect the average rate of supernovae (SNe) to be as high as $\sim 0.1$\,kyr$^{-1}$ as it was estimated by \citet[][]{Mun06} for a rich galactic stellar cluster Westerlund 1 at its current evolution stage. In reality this implies that the system withstands recurrent perturbation and relaxation phases caused by the propagating SNe shock fronts. We performed 3D MHD modeling of SNR evolution inside YMSC in order to study the energy partitioning, thermal and magnetic structure of a cluster right after the SN event. The metal rich matter ejected by SN is expected to change the plasma composition in a cluster for $\sim$ 1000 years and thus may affect the X-ray spectrum of the hot cluster interior. Shocks and magnetic field structure are the key ingredients of non-thermal particle acceleration and radiation models in YMSCs.   

The paper is organized as follows. The 3D MHD model using the PLUTO code which includes the numerical scheme, stellar cluster setup and the supernova initialization are discussed in \S\ref{sec:numsim}.  The simulated plasma density, temperature and magnetic fields distributions in YMSC perturbed by supernova events are presented in \S\ref{sec:res} where 3D rendering images and plane maps both for the central and peripheral locations of SN are considered. The temporal evolution of the SN, energy partitions, statistical distributions of the cluster magnetic fields and SN ejected mass are presented in \S\ref{sec:dis}. In this section we also illustrate the X-ray spectra of the hot optically thin plasma and discuss in brief the non-thermal components in the simulated cluster.     

\section{Numerical simulations}
\label{sec:numsim}

\subsection{Governing equations}
\label{sec:goveq}

The simulations were performed using the well-proven open source code \textmc{pluto} \citep[][]{Mig07,Mig12,Mig18} based on the Godunov method, and created specifically for problems of computational astrophysics. According to our problem, the code integrates the following set of non-ideal magnetohydrodynamic equations:
\begin{gather}
    \frac{\partial\rho}{\partial{t}}+\vect{\nabla}\cdot\left(\rho\vect{u}\right)=0,\label{1}\\
    \frac{\partial\vect{m}}{\partial{t}}+\vect{\nabla}\cdot\left(\vect{mu}-\vect{BB}+\vect{I}p_{\mathrm{tot}}\right)=0,\label{2}\\
    \frac{\partial{E}}{\partial{t}}+\vect{\nabla}\cdot\left[\left(E+p_{\mathrm{tot}}\right)\vect{u}-\vect{B}\left(\vect{u}\cdot\vect{B}\right)\right]=\vect{\nabla}\cdot\vect{F}_{\mathrm{c}}+\Phi\left(T,\rho\right),\label{3}\\
    \frac{\partial\vect{B}}{\partial{t}}+\vect{\nabla}\cdot\left(\vect{uB}-\vect{Bu}\right)=0,\label{4}
\end{gather}
where $\vect{m}=\rho\vect{u}$ represents the momentum density vector of a control volume, $\vect{B}$ is the magnetic field vector, $\vect{I}$ is the identity matrix, $p_{\mathrm{tot}}=p+\vect{B}\cdot\vect{B}/2$ is the total pressure. The total energy density of the systems reads,
\begin{equation}
    E=\frac{p}{\gamma-1}+\frac{\vect{m}\cdot\vect{m}}{2\rho}+\frac{\vect{B}\cdot\vect{B}}{2},
\end{equation}
and the sound speed, $c_{\mathrm{s}}=\sqrt{\gamma{p}/\rho}$, closes the above system of equations, where $\gamma=5/3$ is the adiabatic index. The source term $\Phi\left(T,\rho\right)$ on the right-hand side of the total energy conservation equation represents optically thin radiative losses and heating. The plasma heat flux is determined by the vector $\vect{F}_{\mathrm{c}}$.  

We took into account the gains and losses by optically thin radiative processes following the recipe from \citet[][]{Mey14,Gre19} for the case of photo-ionization equilibrium:
\begin{equation}
    \Phi\left(T,\rho\right)=n_{\mathrm{H}}^2\Gamma\left(T\right)-n_{\mathrm{H}}^2\Lambda\left(T\right),
\end{equation}
where $\Gamma\left(T\right)$ and $\Lambda\left(T\right)$ are the radiative heating and cooling rates, respectively, and $n_{\mathrm{H}}$ is the hydrogen number density. 

The efficiency of thermal conduction in the collisionless turbulent plasma is still under debate. The recent work of \citet[][]{TC22} revealed strong suppression of the thermal conduction in the turbulent, weakly collisional, magnetized plasma by more than two orders of magnitude comparing with the Spitzer's thermal conduction in collisional  plasma. Strong supernova shocks are expected to produce a highly enhanced level of plasma turbulence (at scales well below of that resolved in the MHD simulations) which would suppress the thermal conduction on the time scales studied in the paper. The effect of the \citet[][]{Bal13} solar wind thermal conduction at the appropriate time and length scales on the plasma flows in SSC was studied in Paper~I.

This system of equations is solved by use of the unsplit second-order Runge-Kutta algorithm with linear reconstruction of the variables between adjacent cells together with the combination of HLLD \citep[][]{MK05,Mig07} and HLL \citep[][]{Har83} Riemann solvers. The divergence-free condition for the magnetic field is ensured by the Hyperbolic Divergence Cleaning algorithm \citep[][]{Ded02} in the whole computational domain. The time-marching algorithm is controlled by the standard Courant-Friedrichs-Lewy parameter that we set to $C_{\mathrm{cfl}}=0.2$.

\subsection{Supernova remnant initialization}
\label{sec:init}

For the simulations of the SN event inside the star cluster environment we used a mapping strategy \citep[e.g.][]{Mey15}. Assuming the SNR shock to be isotropic, which is sufficient for our scope, we calculated its early expansion in 1D with high resolution. This simulation accurately tracks the evolution of an SNR starting from the time $t_{0} \approx 10^{-2}$\,yr after the core-collapse up to the time $t_{\mathrm{map}} \approx 30$\,yr as it propagates through the wind of a WR-star progenitor (i.e. SN Type Ib/c). Then, we mapped the obtained 1D solution into the 3D domain, which contains the pre-simulated ISM of an SSC from Paper~I with randomly oriented stellar spins (see it for the details on the interacting winds model), replacing the fixed wind injection region of the progenitor star.

\begin{figure}
    \includegraphics[width=8.5cm]{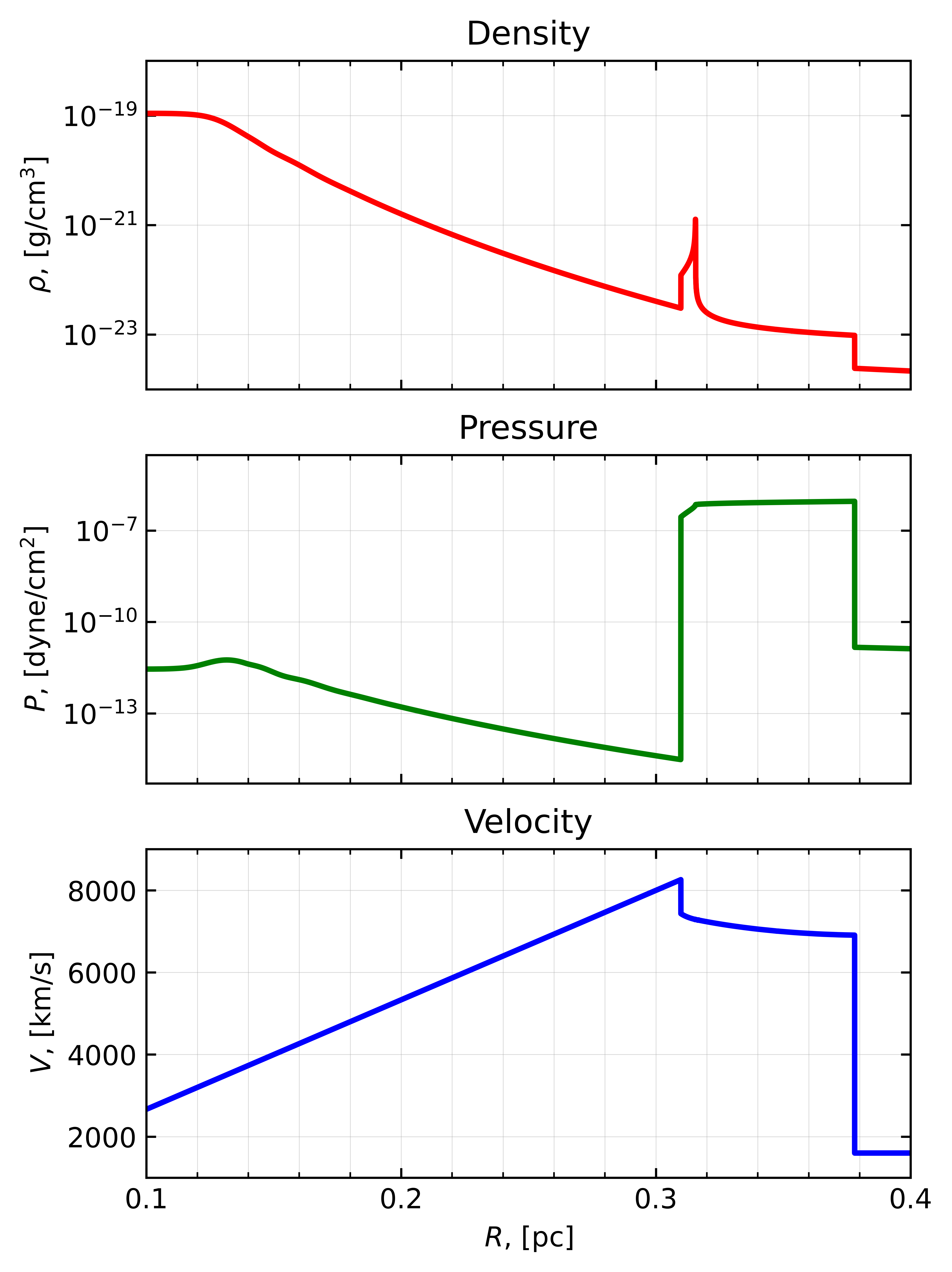}
    \vspace*{-2mm}
    \caption{The obtained 1D SNR solution profiles of density, pressure, and velocity at the moment of mapping $t_{\mathrm{map}}$. Here the forward shock is located at $R=0.38$\,pc and the contact discontinuity between the SN ejecta and progenitor's wind material is right at the density profile spike.}
    \label{fig:SNR}
\end{figure}

For the SNR model we followed a standard procedure \citep[e.g.][]{CL89, TM99, Wha08, Tel13, Pet21} that leads  to the classical self-similar type solution of \citet[][]{Chv82} and \citet{Nad85}, see Fig.~\ref{fig:SNR}. It was assumed that at the time $t_{0}=r_{0}/v_{0}$ shortly after the SN event the ejecta expands freely at the velocity $v_{0}=30000$\kms, and consists of an inner (core) and outer layers with density profiles are being, respectively, uniform and steep:
\begin{equation} \label{rho_steep}
   \rho\left(r\right) =
   \begin{cases}
     F t_{0}^{-3}\hspace{40pt} \mathrm{if}\hspace{5pt}r < r_{\mathrm{c}}, \\
     F t_{0}^{-3}\left(\frac{r}{r_{\mathrm{c}}}\right)^{-n}\hspace{10pt} \mathrm{if}\hspace{5pt}r_{\mathrm{c}} \le r \le r_{0}, \\
     D r^{-2}\hspace{38pt} \mathrm{if}\hspace{5pt}r > r_{0},
   \end{cases}
\end{equation}
where the index $n=9$ corresponds to a Type Ib/c remnant \citep[see][]{Dwa05}, and $D=\dot{M}/4\pi v_{w}$ is a progenitor's wind parameter. Here, $F$ and $v_{\mathrm{c}}=r_{\mathrm{c}}/t_{0}$ are normalization constants that must be determined from the required ejecta mass $M_{\mathrm{ej}}=10$\Msun and SN event energy $E_{\mathrm{ej}}=10^{51}$\,erg. In general, in order to find these constants one has to follow the two-step numerical procedure described in \citet[][]{Wha08}, or, if the inequality $r_{\mathrm{c}} \ll r_{0}$ deliberately holds, use the following analytic expressions:
\begin{gather}
   F = \frac{1}{4 \pi n}\frac{\left[3\left(n-3\right)M_{\mathrm{ej}}\right]^{5/2}}{\left[10\left(n-5\right)E_{\mathrm{ej}}\right]^{3/2}}, \\
   v_{\mathrm{c}} = \sqrt{\frac{10\left(n-5\right)E_{\mathrm{ej}}}{3\left(n-3\right)M_{\mathrm{ej}}}}.
\end{gather}

In our case the circumstellar environment around the expanding SNR is a free WR wind that inherits its parameters from Paper~I, hence: $\dot{M}=6.50 \times 10^{-5}$\MsunYr, and $v_{\mathrm{w}}=1600$\kms. We did not consider the magnetic field in this 1D SNR solution, as (1) it is not dynamically important at the earliest free-expansion SNR phase \citep[see][]{Das22}, (2) it is assumed to contribute a quantitatively negligible fraction to the interstellar magnetic fields that will be compressed by the SNR forward shock.

We used a simplified description of the wind-driven circumstellar medium around the supernova progenitor star which is specific for a SN in a YMSC. More complex model could account for evolutionary changes in the stellar mass loss rate \citep[see e.g.][]{2022Galax..10...37D} which is important for isolated massive stars. In the case of SN in a YMSC of a few Myr age, which we are discussing here, the surroundings of the progenitor star are rather determined by the colliding winds of massive stars on sub-parsec scale sizes. Also, the possible previous SNe could sweep out the matter from the cluster and thus erase the stellar evolution history to a significant extent.

Once the initial 1D SNR solution was mapped into 3D MHD domain the \textmc{pluto} code solved the equations (\ref{1}--\ref{4}) in Cartesian grid $(x,y,z)$. The computational domain was extended in the intervals of $[-2.16;2.16]$\,pc in all directions and covered with a uniform grid of $540^{3}$ and $270^{3}$ cells for the first 1500 yr and the last 6000 yr (down-scaled relaxation part, see $\S\,$\ref{relax}) of integration time, respectively. At the domain borders we used a modified 'free outflow' boundary condition that prohibits any possible backflow of the gas.

We performed two simulations for the two cases of the SNR progenitor position inside the cluster core: near the cluster center, $r_{\mathrm{dist}}\approx0.6$\,pc ('central'), and at the cluster periphery, $r_{\mathrm{dist}}\approx2$\,pc ('peripheral'). In the latter case it was expected that the SNR shock propagating through the whole cluster volume would sweep and compress more material flowing both towards (in the first half of the cluster volume) and away (in the second half) from the SNR forward shock, possibly resulting in the more effective Axford-Cranfill-type magnetic field amplification.

The ejecta material was traced using the scalar marker $Q$ which was described by the linear advection equation:
\begin{equation}
    \frac{\partial\left(\rho{Q}\right)}{\partial{t}}+\vect{\nabla}\cdot\left(\vect{v}\rho{Q}\right)=0.
\end{equation}
This marker was passively advected with the fluid, allowing us to distinguish between the ejecta and interstellar material. We set the value $Q(\vect{r})$ to be negative for the ejecta and positive for the SWs and interstellar material, here $\vect{r}$ is the vector position of a given cell in the simulation domain.

\section{Results}\label{sec:res}

Here we present the results of two simulations from $t_{\mathrm{snr}}=100$\,yr up to $t_{\mathrm{snr}}=500$ and $t_{\mathrm{snr}}=700$\,yr for the centrally located and peripheral SN events inside the YMSC core, respectively. These moments in time track the expansion of the SNR until the moment when its forward shock reaches the computational domain borders. We use two forms of presentation: 2D maps (cross-sections of a 3D domain) and 3D volumetric renders of scalar and vector fields. For both simulations the integration times up to $t_{\mathrm{snr}}=7500$\,yr are thoroughly analysed in \S\ref{sec:dis}.

\begin{figure*}
	\includegraphics[scale=0.25]{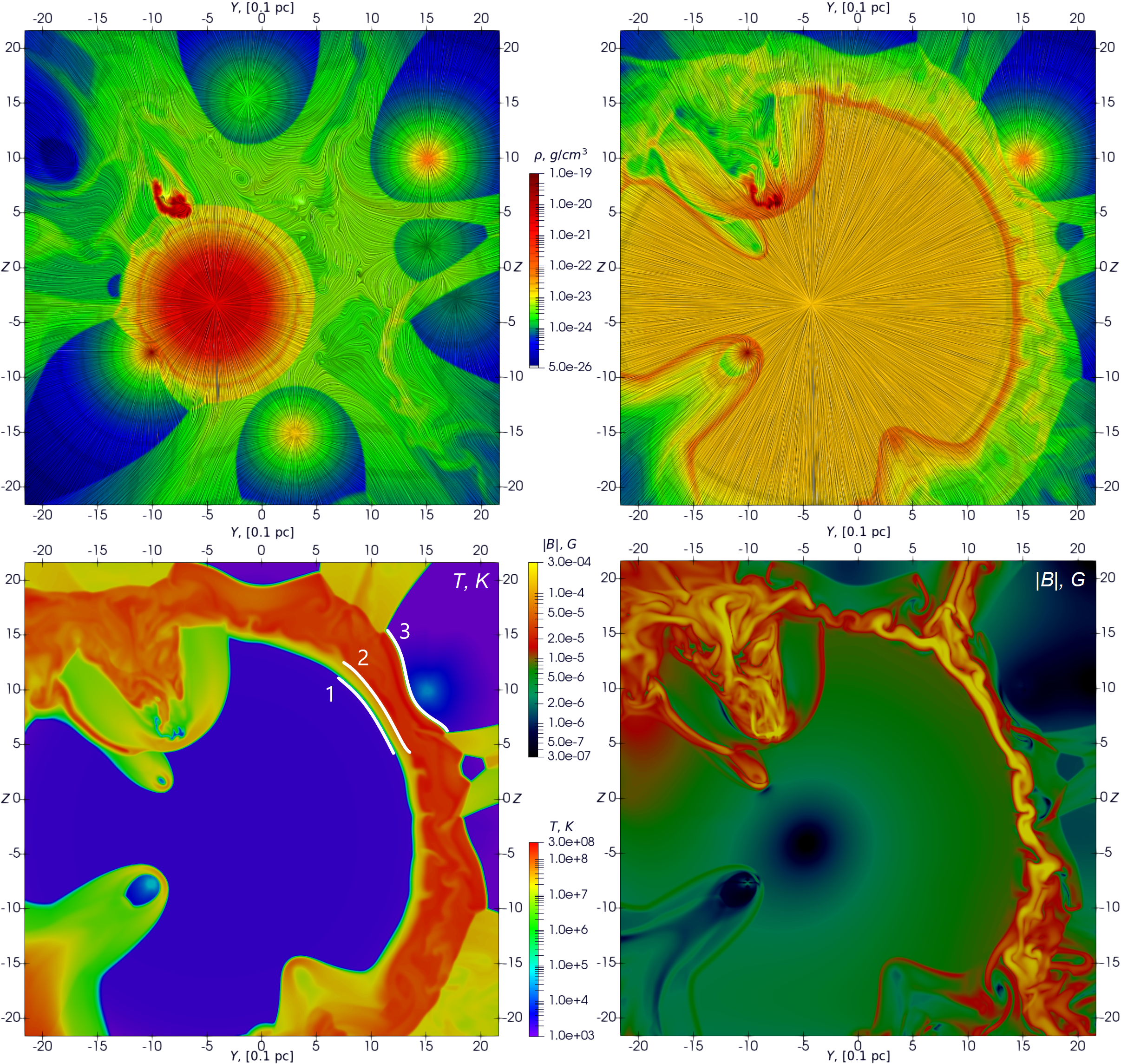}
    \caption{The evolution of the SNR placed near the cluster center. Top: density maps and flow streamlines (dark translucent lines) at $t_{\mathrm{snr}}=100$ and $t_{\mathrm{snr}}=500$\,yr, left to right. Bottom: temperature and magnetic field magnitude maps at $t_{\rm {snr}} = 500$\,yr. The white lines 1, 2 and 3 mark the positions of the reversed shock, contact discontinuity and forward shock, respectively. In the top left corner the dense CSG-wind shell has formed a wide bow shock with a magnetized tail. The maps are captured in $zy$-plane at $x=-2.7$ in units of [0.1 pc].}
    \label{fig:central}
\end{figure*}
\begin{figure*}
	\includegraphics[scale=0.25]{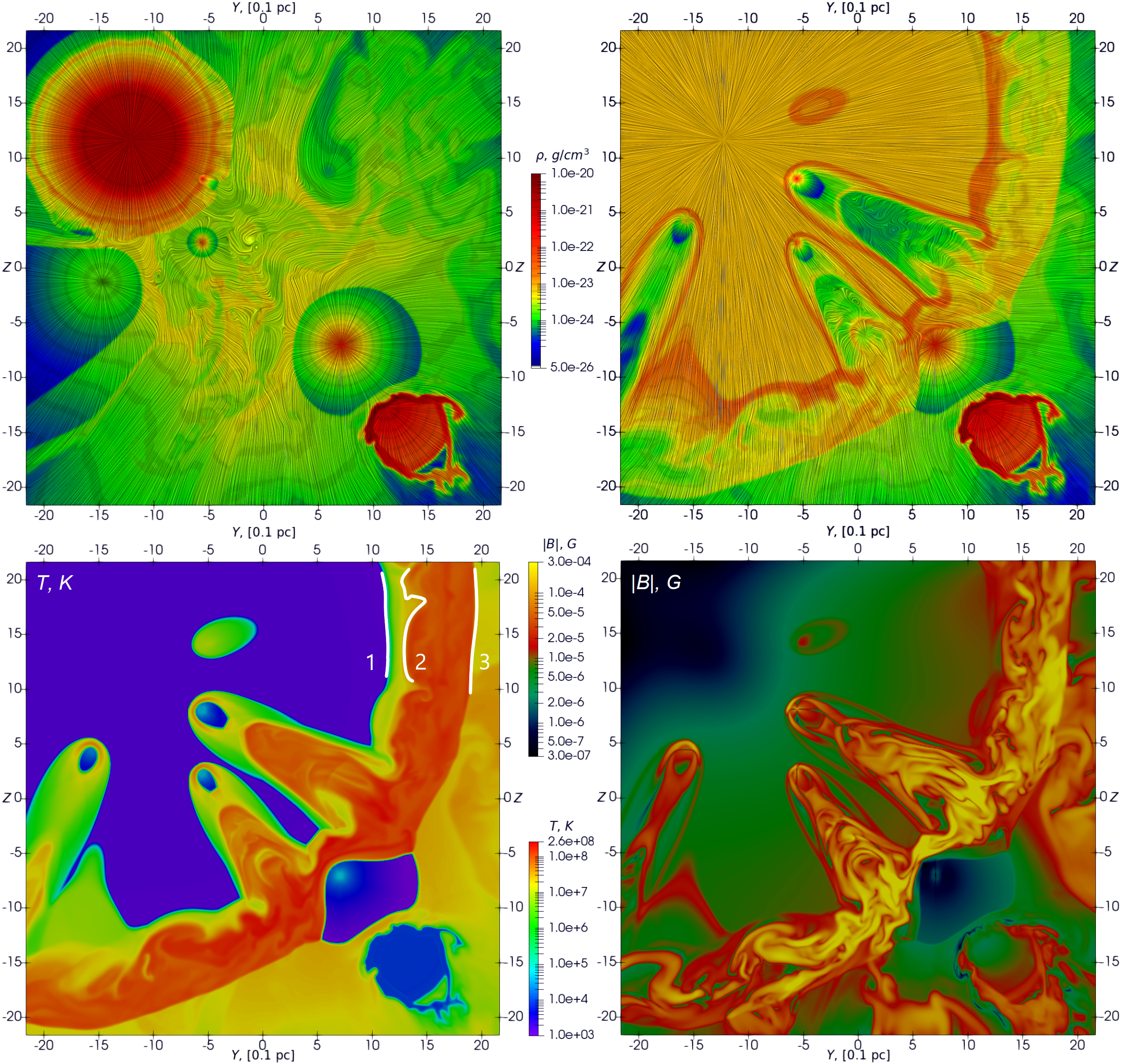}
    \caption{The evolution of the SNR placed at the cluster periphery. Top: density maps and flow streamlines (dark translucent lines) at $t_{\mathrm{snr}}=100$ and $t_{\mathrm{snr}}=700$\,yr, left to right. Bottom: temperature and magnetic field magnitude maps at $t_{\rm {snr}} = 700$\,yr. The white lines 1, 2 and 3 mark the positions of the reversed shock, RT-unstable contact discontinuity and forward shock, respectively. At the bottom right corner there is a cold and dense CSG-wind shell. The maps are captured in $zy$-plane at $x=-10.7$ in units of [0.1 pc].}
    \label{fig:perif}
\end{figure*}

\begin{figure}
        \centering
	\includegraphics[width=8.25cm]{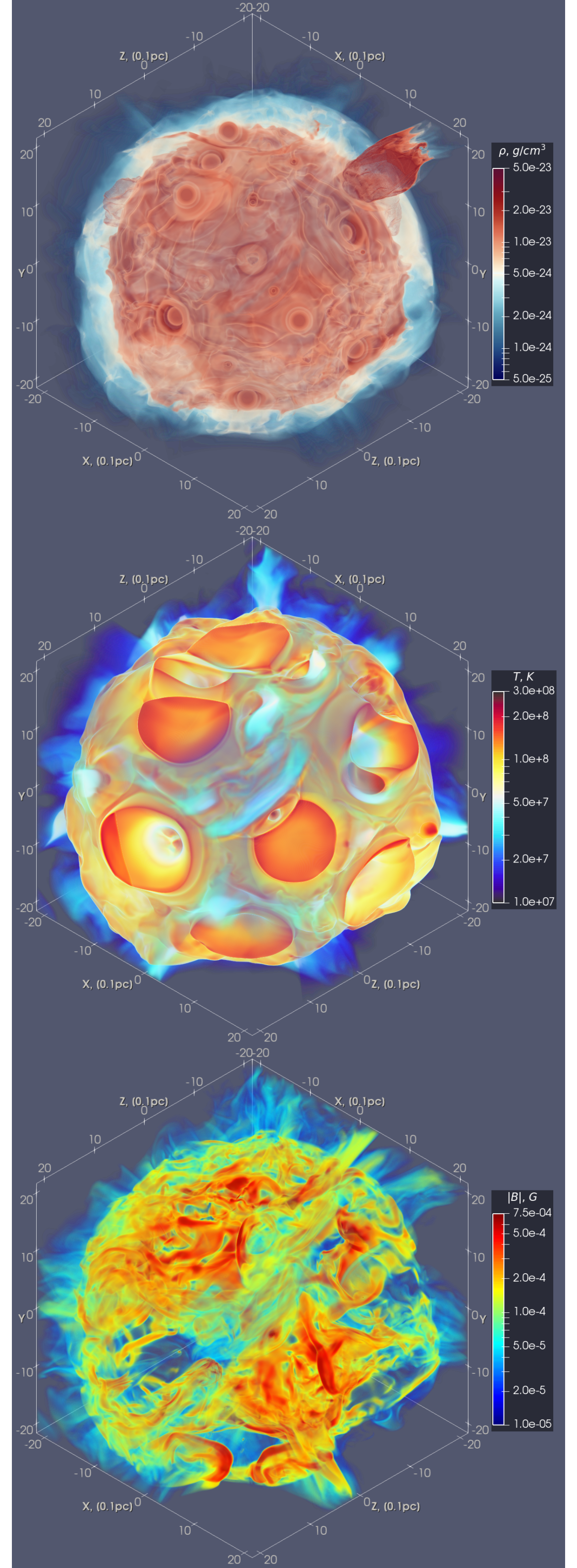}
	\vspace*{-0.5mm}
        \caption{The density (top), temperature (middle) and magnetic field (bottom) distributions at the surface of the cluster-centered SNR shell shown as 3D volumetric renders, $t_{\mathrm{snr}}=500$\,yr. The red bulb in the top figure is a CSG-wind shell. The orange 'dents' in the middle figure are the SWs impacting the SNR.}
        \vspace*{-5mm}
        \label{fig:3D}
\end{figure}

\subsection{Overall dynamics and structure of flows}

We have obtained a 3D MHD data on the passage of the SNR shock wave through the Wd1-like YMSC core at a resolution of $8\times10^{-3}$\pccell over the dynamical time $\tau_{\mathrm{dyn}}=l/v_{\mathrm{prop}}\sim1200$\,yr, where $v_{\mathrm{prop}} \sim 3000$\kms is the SNR shock propagation speed and $l\approx4$\,pc is the length scale. The results are presented in 2D maps of density, temperature, and magnetic field amplitude for two cases of SN location: in the center of the cluster core (see Fig.~\ref{fig:central}) and on its periphery (see Fig.~\ref{fig:perif}). The figures show cross-sections perpendicular to the $x$-axis passing through the points where the SN progenitors were located. Flow streamlines were plotted over the density maps using velocity vector field data. Since the initialized remnant is spherically symmetric, the morphology of the expanded SNR shell depends on the non-uniform and intermittent collective flows inside the simulated cluster. Therefore, it is important to note that different cross-sections may reveal slightly different structures of the SNR shell in detail.

The local structure and morphology of the expanding SNR shell are determined by the winds of neighboring stars. The forward shock crushes through the SWs within the cluster core, creating numerous bow shocks with various geometries depending on the wind type. The size and geometry of a bow shock depend on the balance between the ram pressure of the SNR and the kinetic power of the wind \citep[see][]{Wil96}. The widest and thickest bow shocks are formed around cold and dense CSG-winds, as shown in Fig.~\ref{fig:central}. In the areas of impact, SWs form cometary structures that penetrate through the supernova remnant shell and leave heated up 'dents' on its surface, see Fig.~\ref{fig:3D}. Magnetic fields are amplified to approximately 100 $\mu$G at the bow shocks and tails of these cometary structures. The overall spherical structure of the remnant is preserved during expansion. When the SNR shell leaves the volume of the domain, the SWs relax to their roughly unperturbed sizes in $\sim3000$ yr, as shown in Fig.~\ref{fig:relax}.

\subsection{The SNR expansion}

Just a hundred years after the SN event one could clearly see the remnant with nearly undistorted initial structure, see Figs.~\ref{fig:central}-\ref{fig:perif}. At this stage the SNR expands freely and it has a typical structure with two shock waves (reverse and forward) and a contact discontinuity (CD) \citep[see e.g.][]{TM99}. 

The latter can be easily identified by the Rayleigh-Taylor (RT) instability modes at the surface of the inner dense ($n\sim30$\,cm$^{-3}$) and thin ($\sim0.1$ pc) shell shown at the top of Fig.~\ref{fig:3D}. In the immediate vicinity of the CD, there is a reverse shock that heats a thick but dense layer of SN ejecta. The lack of a reverse shock moving towards the center, as explained in \citep[see][]{Pet21}, can be attributed to the high amount of ejected mass compared to the mass of the cluster diffuse gas and the short time scales. As a result, no extensive interior heating is observed, and most of the SNR volume cools adiabatically. However, this shock effectively heats the inner dense SNR shell to the temperatures $\sim10^{7}$ K. On the other hand, the forward shock is responsible for heating the swept-up interstellar medium plasma to temperatures up to a few $10^{8}$ K, see the temperature data in Fig.~\ref{fig:3D}. It is also responsible for effective amplification of interstellar magnetic fields well above 200 $\mu$G in some filaments and thin isolated regions, see the bottom of Fig.~\ref{fig:3D}. This amplified field is probably carried into the bow shock tails, this is clearly seen in Fig.~\ref{fig:perif}. The thickness of the shocked interstellar gas layer is $\sim0.5$ pc with the density $n\sim4$\,cm$^{-3}$.

Some minor differences in thickness of the shocked gas layer and dense RT-unstable shell are the consequence of different initial positions of the SNRs. No significant quantitative differences in terms of densities, temperatures, and magnetic field strength are detected for the case of peripheral SN over the central one.

\section{Discussion}\label{sec:dis}

\begin{figure*}
    \hbox{\hspace{7.5mm}\includegraphics[scale=0.335]{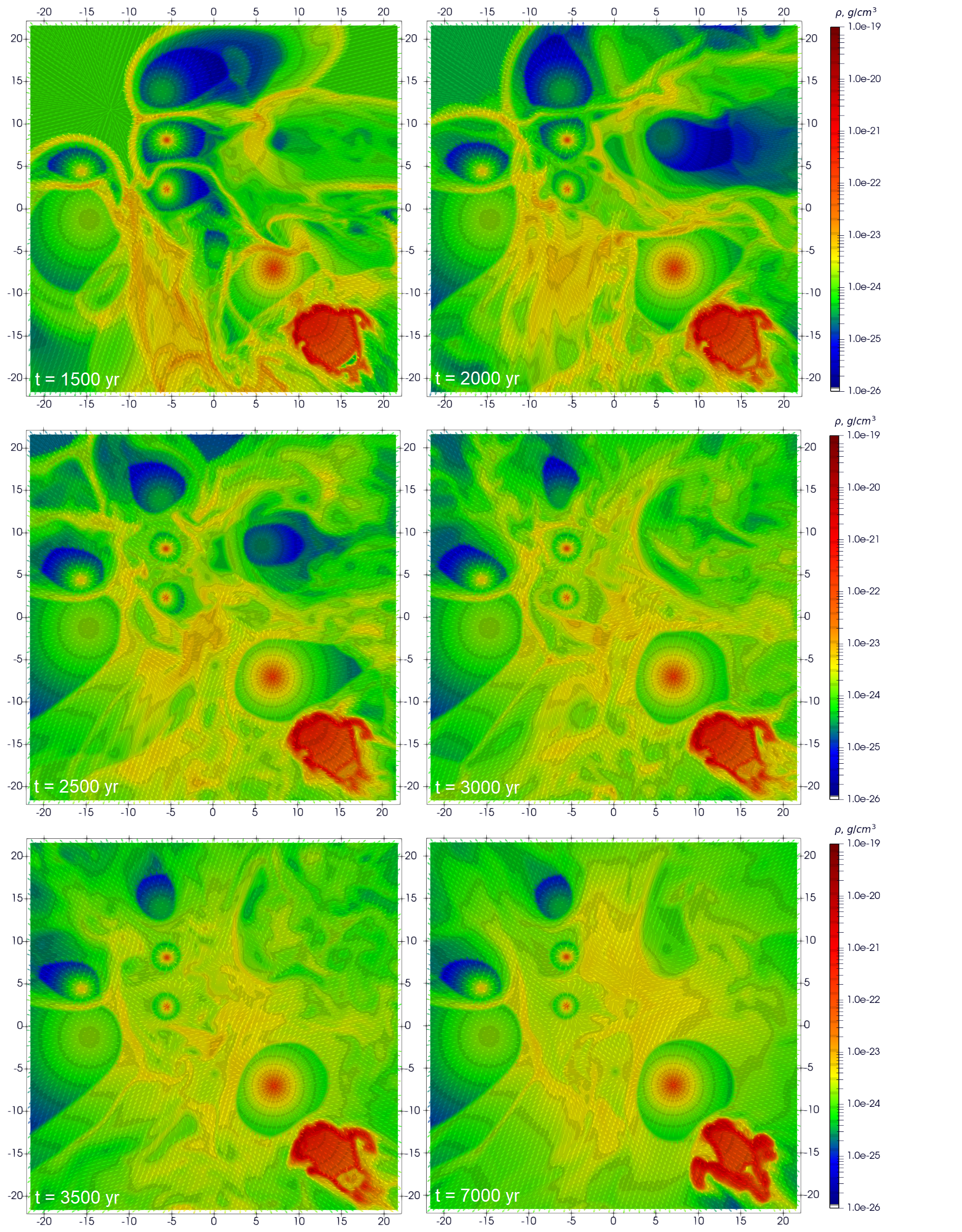}}
    \vspace*{-1mm}
    \caption{Relaxation dynamics after the peripheral SN event. At the time of 3500 yr the overall flow geometry is recovered, and after the next 2500 yr the small-scale density perturbations are smoothed. At the lower right corner of each frame one can see a dynamically stable dense shell which is formed by a CSG-wind. The cross-section here is the same as the one given in Fig.~\ref{fig:perif} ($zy$-plane), so the axis tick marks are also given in units of [0.1 pc].}
    \vspace*{-5mm}
    \label{fig:relax}
\end{figure*}

\begin{figure*}
    \includegraphics[width=8.5cm]{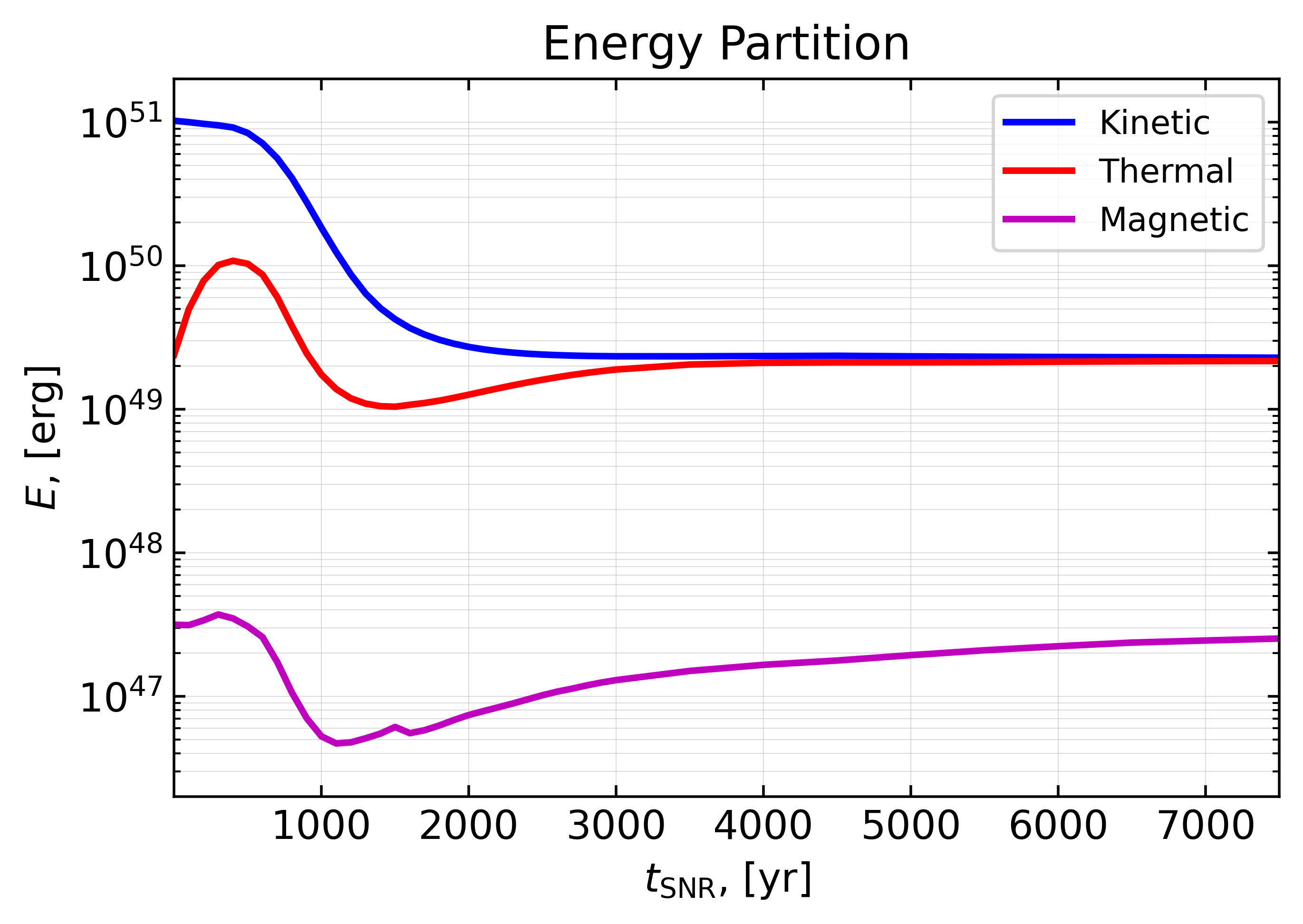}
    \includegraphics[width=8.5cm]{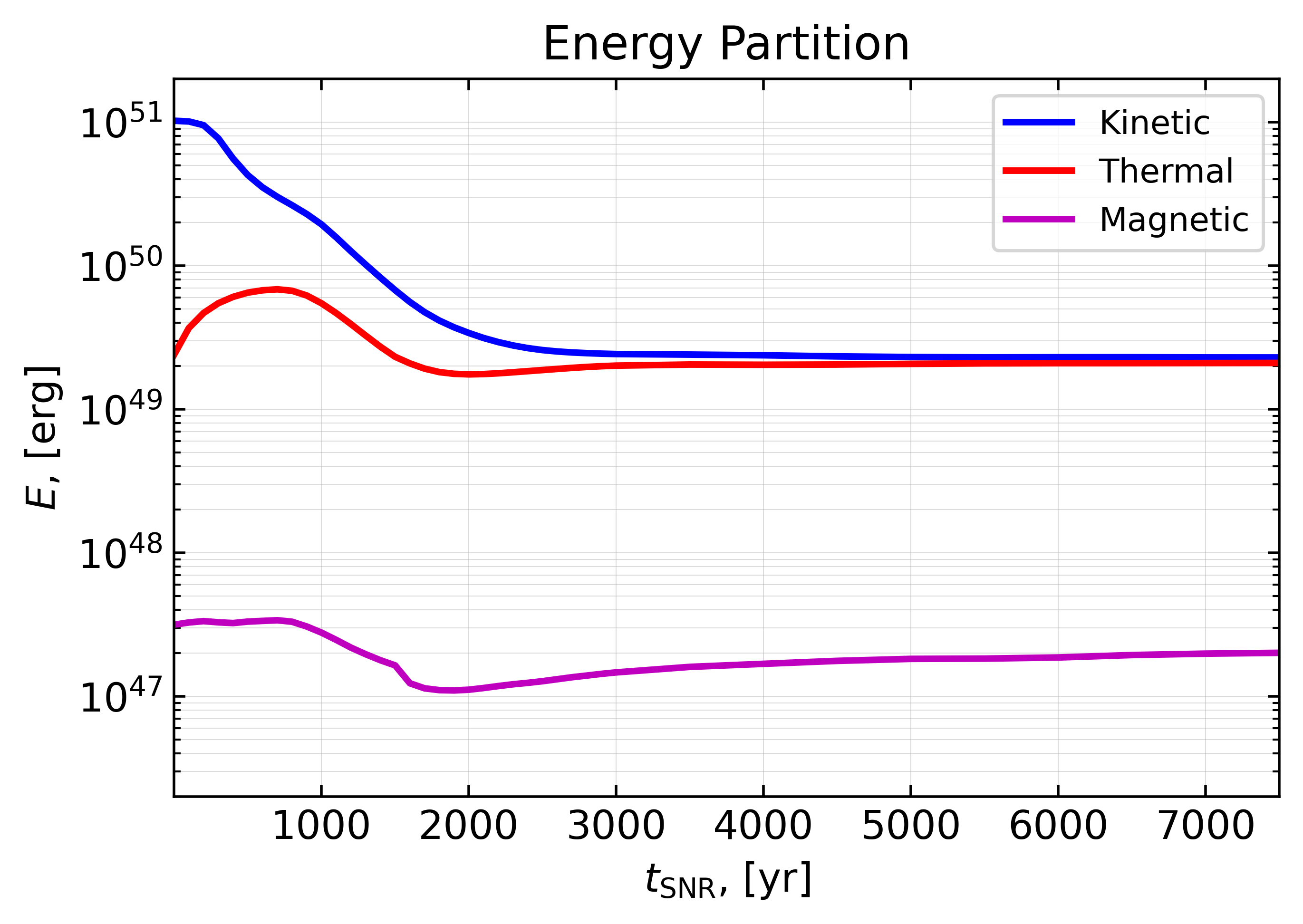}
    \vspace*{-2mm}
    \caption{Total kinetic, thermal, and magnetic energy pools as functions of time $t_{\mathrm{snr}}$. The left picture stands for the central and the right one stands for the peripheral SN event. Due to the resolution down-scaling the magnetic energy curves have a small drop at $t_{\mathrm{snr}}=1500$\,yr.}
    \label{fig:enrg}
\end{figure*}

\subsection{Intracluster medium relaxation}\label{relax}

For a stellar cluster similar to Wd1 the most massive stars are likely to produce the first SN events when the cluster is $\sim 3$ Myr old, and then, following the estimation by \cite{Mun06}, one should expect $\sim 1$ SN every 7-13 kyr for the time interval of duration $\gtrsim 1$ Myr. It is important therefore to derive the relaxation times of the intracluster plasma after the SN event. These times are somewhat different for the dynamics, energy partition, and ejected material admixture. The cluster crossing time by the SN shock is $\lesssim 1000$ yr depending on the progenitor star position. The simulation presented above shows that the violent disturbance of the intracluster medium produced by the core-collapse SN shock flow relaxes to the quasi-stationary state 4000-5000 yr after the SN event. This is consistent with what we see in Fig.~\ref{fig:relax}, where the general structure of flows is recovered at $\sim 3500$ yr and during the next several thousand years the small-scale density perturbations are being suppressed. In terms of kinetic and thermal energy partitioning, as it is apparent in Fig.~\ref{fig:enrg}, it takes $\sim 4000$ yr for the system to reach the relaxed state. Yet one should note that in the case of the peripheral SN event the sub-dominant magnetic energy part, while slowly reaching some plateau at 7000 yr, still struggles to recover back to its base level at $\approx 3 \times 10^{47}$ erg. This could be both the effect of resolution down-scaling after $t_{\rm snr}=1500$ yr and the consequence of slow magnetic field reorganisation. If we track the ejecta admixture (see Fig.~\ref{fig:mass_tr}), the relaxation takes place at $\sim 4000$ yr, when highly diluted leftovers of the ejecta finally leave the domain volume, regardless of the SN event initial position.

The energy partition and total mass of the gas presented in Figs.~\ref{fig:enrg} and \ref{fig:mass} as functions of time were integrated explicitly:
\begin{equation}
A\left(t\right)=\sum_{\substack{i,j,k=1}}^{n_{\rm{cell}}}a_{ijk}\left(t\right)V_{\rm{cell}},
\end{equation}
where $a_{ijk}(t)$ represents the scalar density field, $V_{\rm{cell}}$ is the domain control volume, and $n_{\rm{cell}}=270$. For the mass loss rate estimate we calculated the integral of the flow $\rho_{ijk}(t)\vect{v}_{ijk}(t)$ through the faces of the cubic computational domain.

\subsection{Magnetic field evolution}\label{mf}

Magnetic fields in the simulated core of a YMSC are highly intermittent and span a broad range of magnitudes from a few to hundreds of $\mu$G with various volume filling factors \citep[cf.][]{Bad22}. The energetic SN event severely disturbs the quasi-stationary magnetic field configuration as well as the other important characteristics inside the cluster (see \S\ref{relax}). The 3D render of the magnetic field structure for $t=500$ yr after the centrally located SN event is shown in Fig.~\ref{fig:3D}. 

\begin{figure*}
    \includegraphics[width=8.415cm]{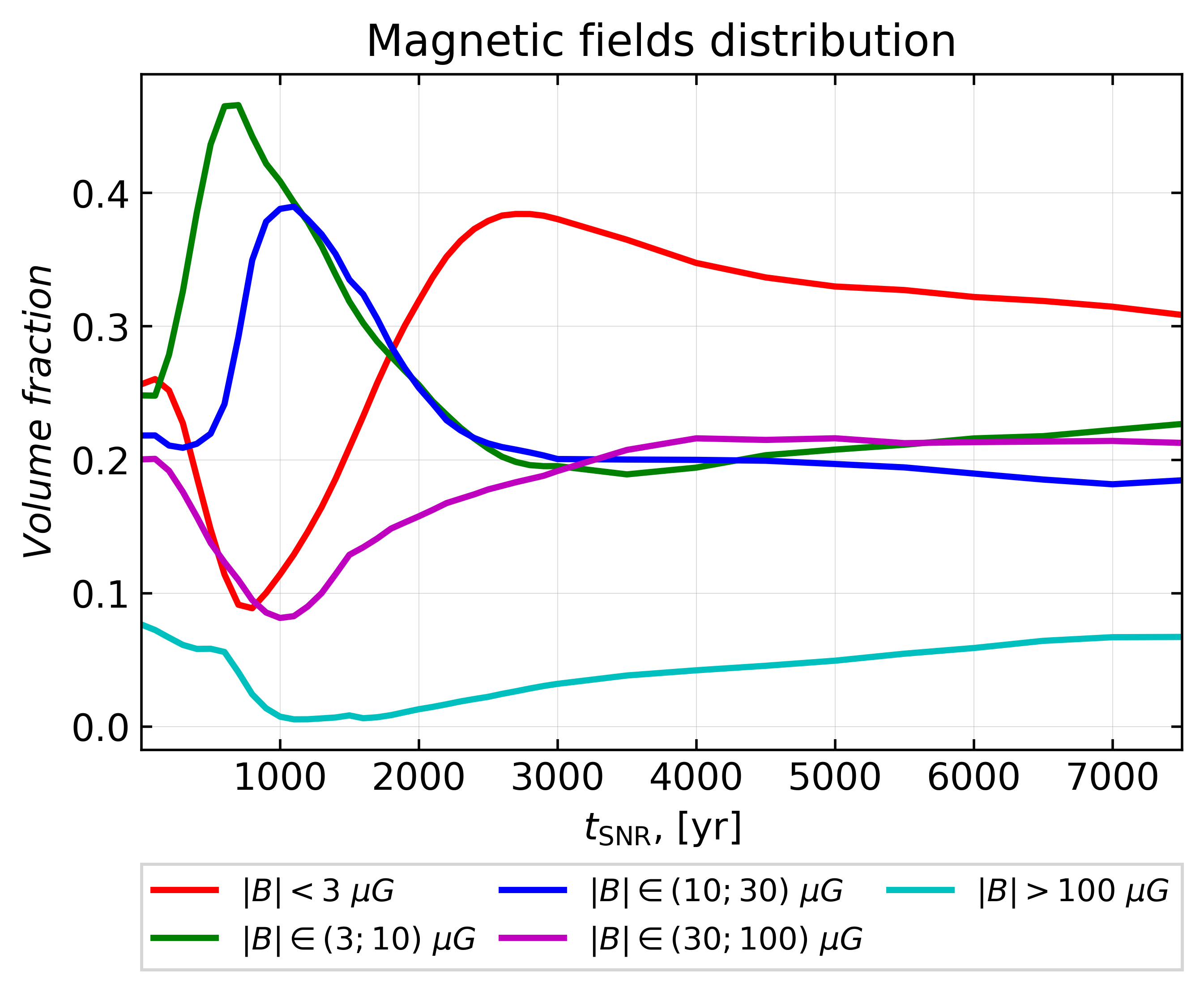}
    \includegraphics[width=8.585cm]{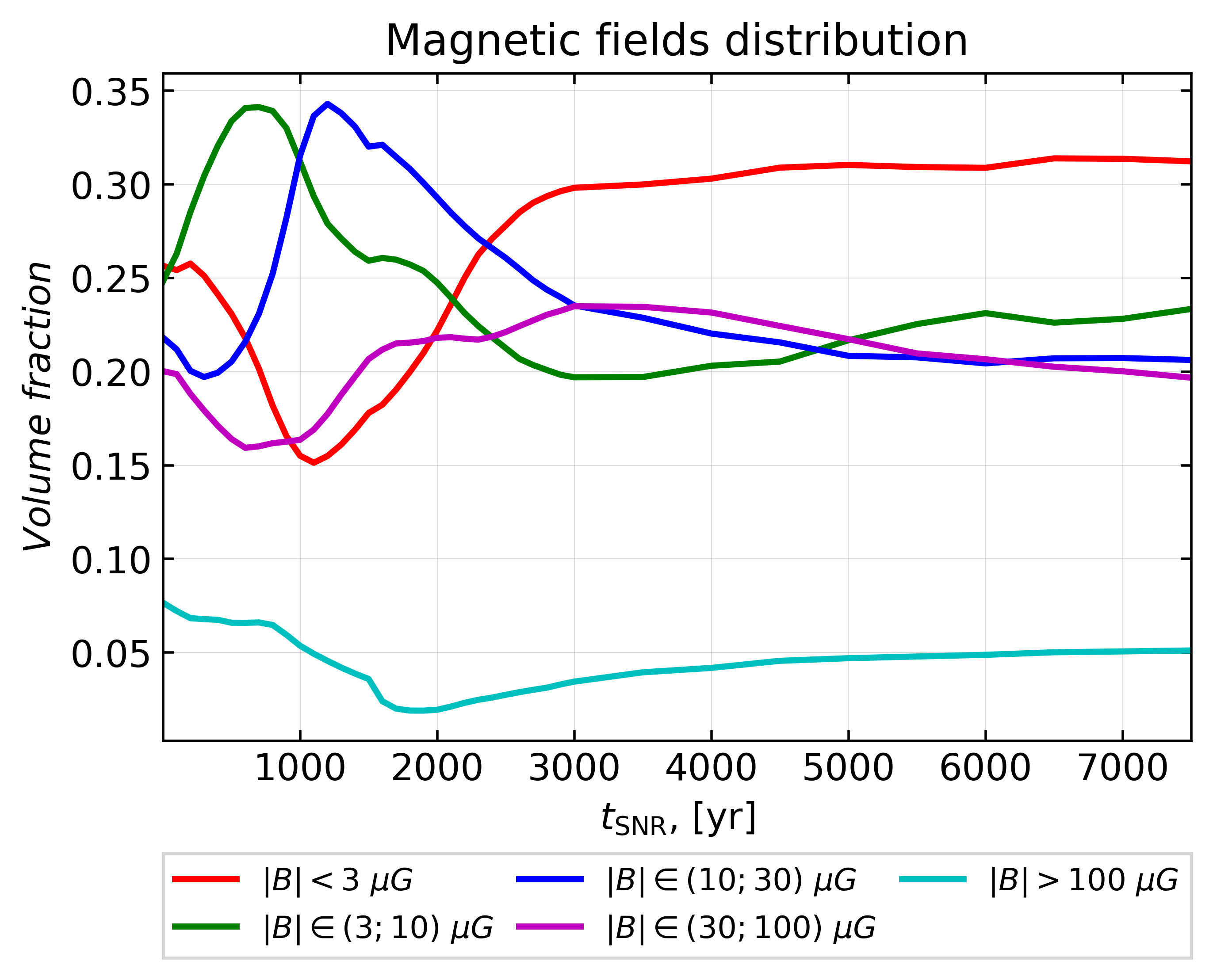}
    \vspace*{-2mm}
    \caption{Distribution of the magnetic field in different bands of absolute value. The left picture stands for the case of centrally located SN and the right one stands for the peripheral SN event. It is clearly visible that during the propagation of the SNR shock through the cluster core volume the range of $\lvert B \rvert\in(3;30)$\muG becomes dominant in terms of volume.}
    \label{fig:B_dis}
\end{figure*}

The magnetic energy of the simulated cluster core is dominated by the magnetic fields of amplitudes $\gtrsim100$ $\mu$G with a few per cent volumetric filling factor. This is consistent with the results of \citet{inoue_MFA09} 2D simulations of a young supernova remnant in a turbulent environment. One can see that the magnetic field magnitude reaches the values $\lesssim750$~$\mu$G in some highly compressed regions of the filamentary structures, see Fig.~\ref{fig:3D}. The evolution of the magnetic structures over time can be observed in Fig.~\ref{fig:B_dis}. It shows the changes in the filling factors of magnetic fields of different magnitudes for both the central and peripheral supernova events. The temporal evolution differs between central and peripheral supernovae, but in both cases, a quasi-stationary distribution is reached after $\sim4000$ yr. 

The decrease in volumetric filling factors observed in high amplitude magnetic field bands can be attributed to the crushing and sweeping of the initial magnetic field configuration which was previously formed by the colliding winds of massive stars. The strong forward shock driven by the SNR travels through highly inhomogeneous cluster medium. It compresses the interstellar magnetic field immediately behind the collisionless shock and possibly amplifies the fluctuating fields downstream through dynamo-like effects. The thin shell of high magnetic field accompanies the SNR forward shock and the downstream flows, overrunning powerful stellar winds as depicted by magnetic field data in Figs.~\ref{fig:central}-\ref{fig:perif} and \ref{fig:3D}.

\begin{figure*}
    \includegraphics[width=8.5cm]{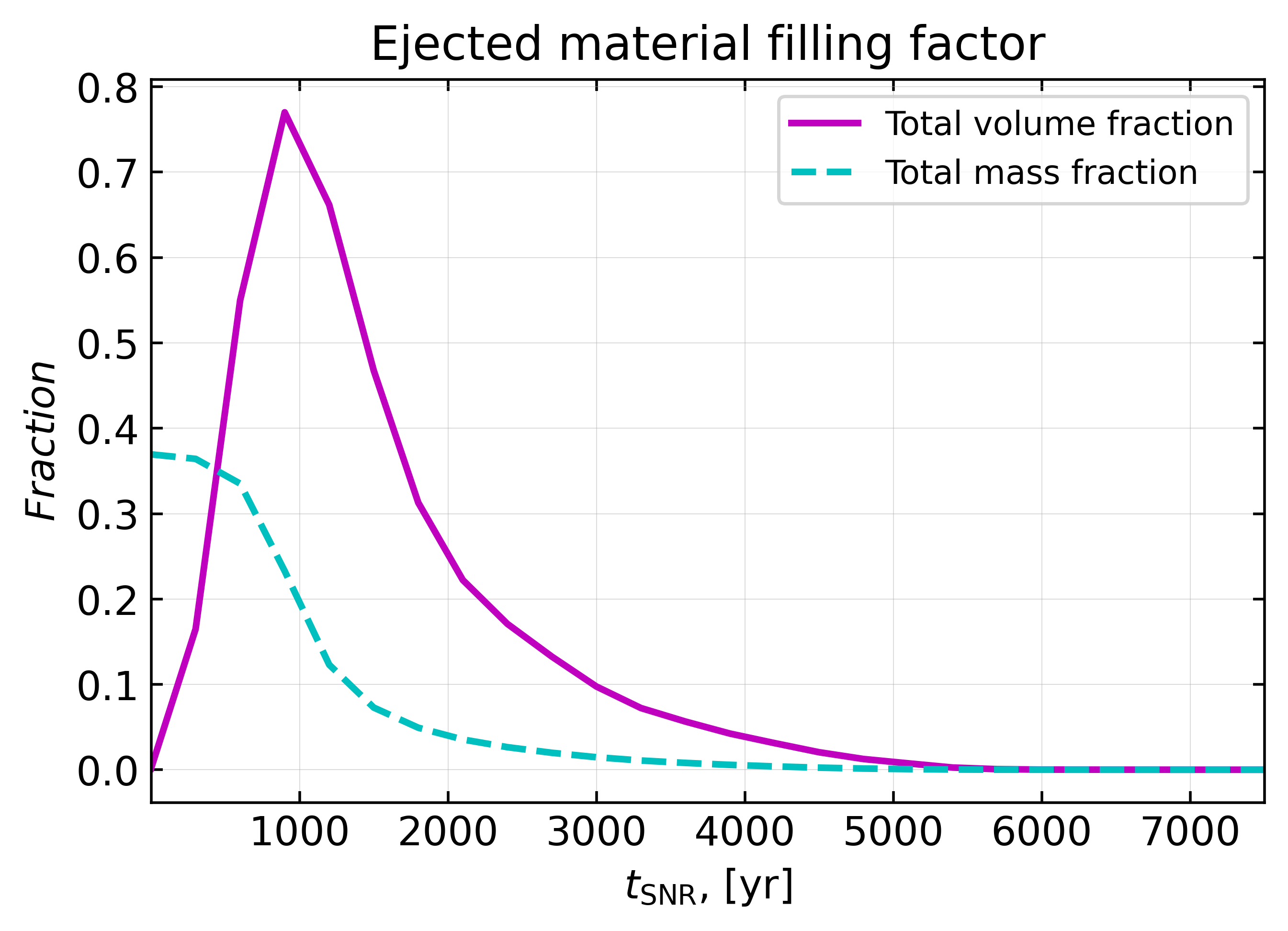}
    \includegraphics[width=8.5cm]{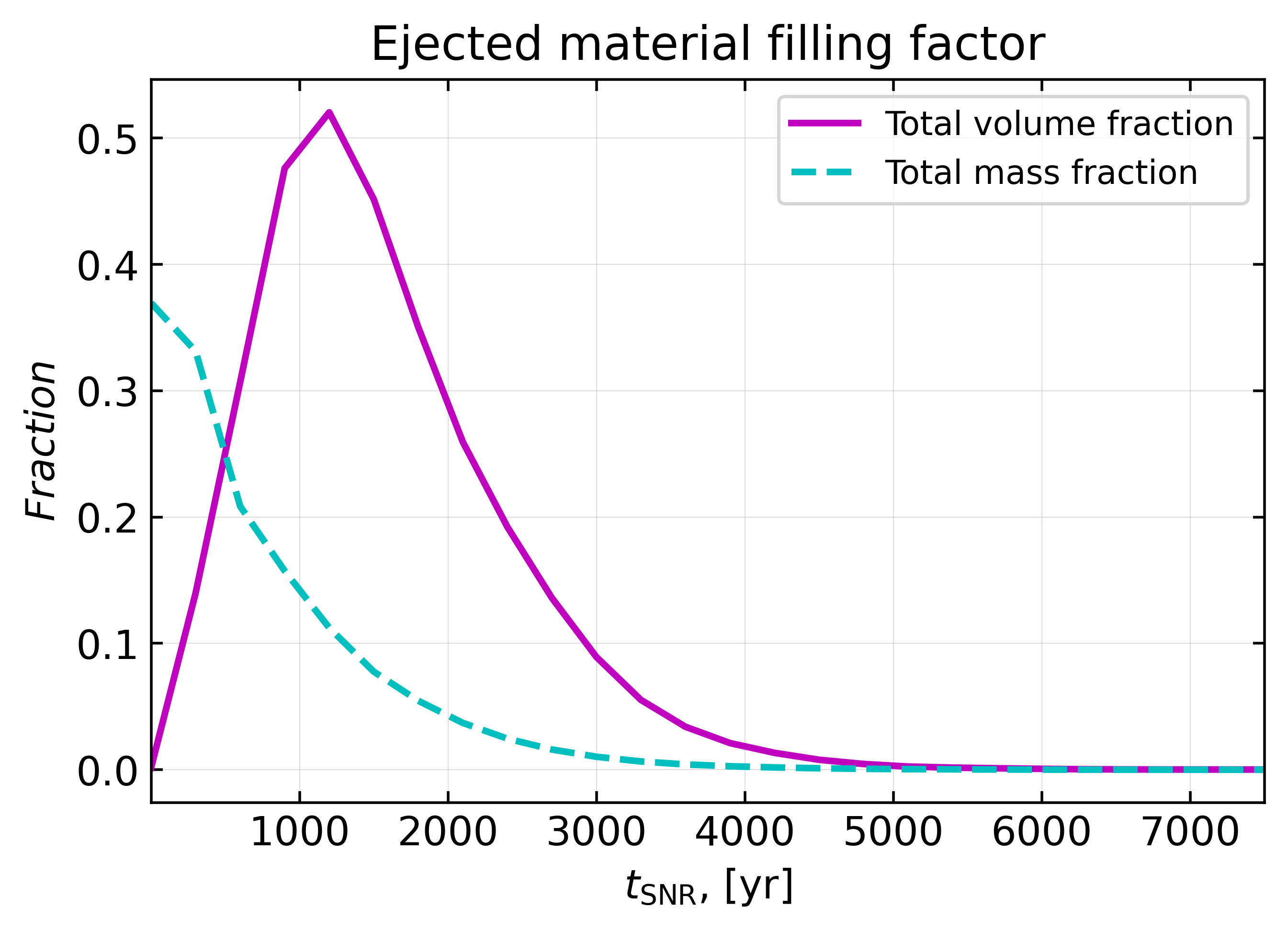}
    \vspace*{-2mm}
    \caption{Magenta line shows the part of the cluster core (domain) volume filled with the SN ejecta. Cyan dashed line is a ratio of the ejected mass to the total diffuse mass contained inside the cluster core volume. The left picture stands for the central SN and the right one stands for the peripheral SN event.}
    \label{fig:mass_tr}
\end{figure*}

The total magnetic energy of the cluster is $\sim10^{47}$ erg and exhibits some apparent temporal evolution especially during the central SN event (see Fig.~\ref{fig:enrg}). The magnetic energy has a prominent minimum that occurs when the SNR shock front exits the domain volume and lasts until the fast colliding SWs start restoring the magnetic fields in the cluster core, before the next possible SN event. It is evident that during the propagation of the SNR shock through the cluster core volume, the range of magnetic field magnitudes $|\vect{B}|$ between 3 and 30 $\mu$G becomes spatially dominant. The total energy in magnetic fields is mainly contributed by fields with magnitudes well above 30 $\mu$G, with a significant contribution from regions with magnitudes higher than 100 $\mu$G, as seen in Figs.~\ref{fig:enrg}-\ref{fig:B_dis}. The presence of these high magnetic fields can be tested by searching for synchrotron X-rays from multi-TeV leptons, which are expected to be present in TeV sources like Westerlund 1.

In Fig.~\ref{fig:B_dis} the volumetric filling factors for the magnetic fields in range $b=\left(B_{\rm min};B_{\rm max}\right)$ were calculated as follows:
\begin{equation}\label{eq:filfac}
    F_{b}(t)=\frac{1}{V}\sum_{\substack{i,j,k}}V_{\rm cell},\hspace{10pt}(i,j,k):\hspace{2pt}|\vect{B}_{ijk}(t)|\in b,
\end{equation}
where $V$ stands for the total domain volume, and $|\vect{B}_{ijk}(t)|$ is the magnetic field magnitude.

\subsection{Ejecta material transport}\label{ej}

In this paper we modelled the ejecta relaxation in the compact cluster with multiple powerful SWs at a few kyr time-scale. This process is relevant to the YMSCs evolution stage dominated by the most massive stars. While the radiative losses of the hot gas in the cluster do not affect seriously the plasma dynamics in any case, the account of the metal enhancement from the SN ejecta material could be important for the X-ray diffuse emission spectra. 

\begin{figure*}
    \includegraphics[width=8.5cm]{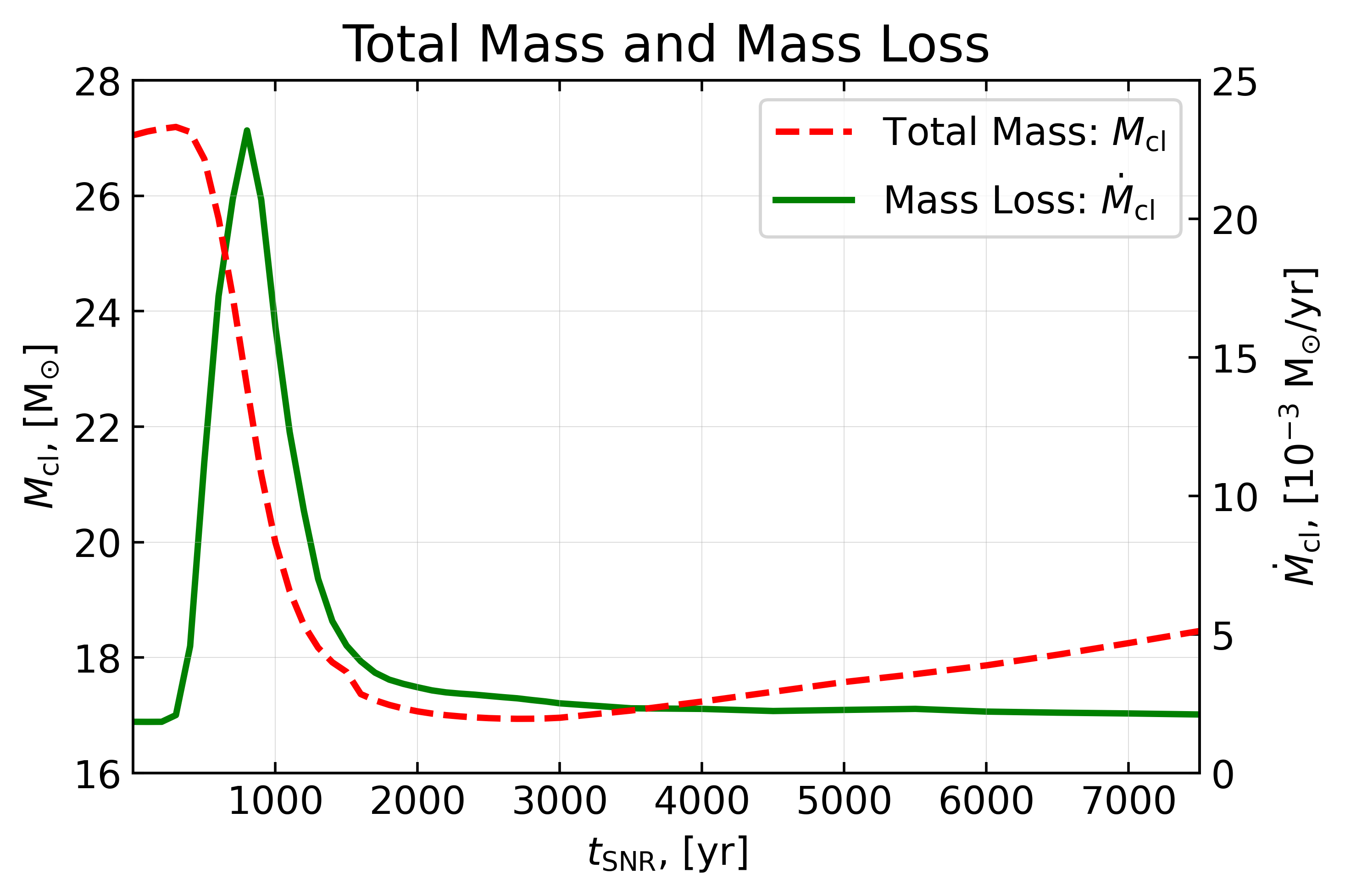}
    \includegraphics[width=8.5cm]{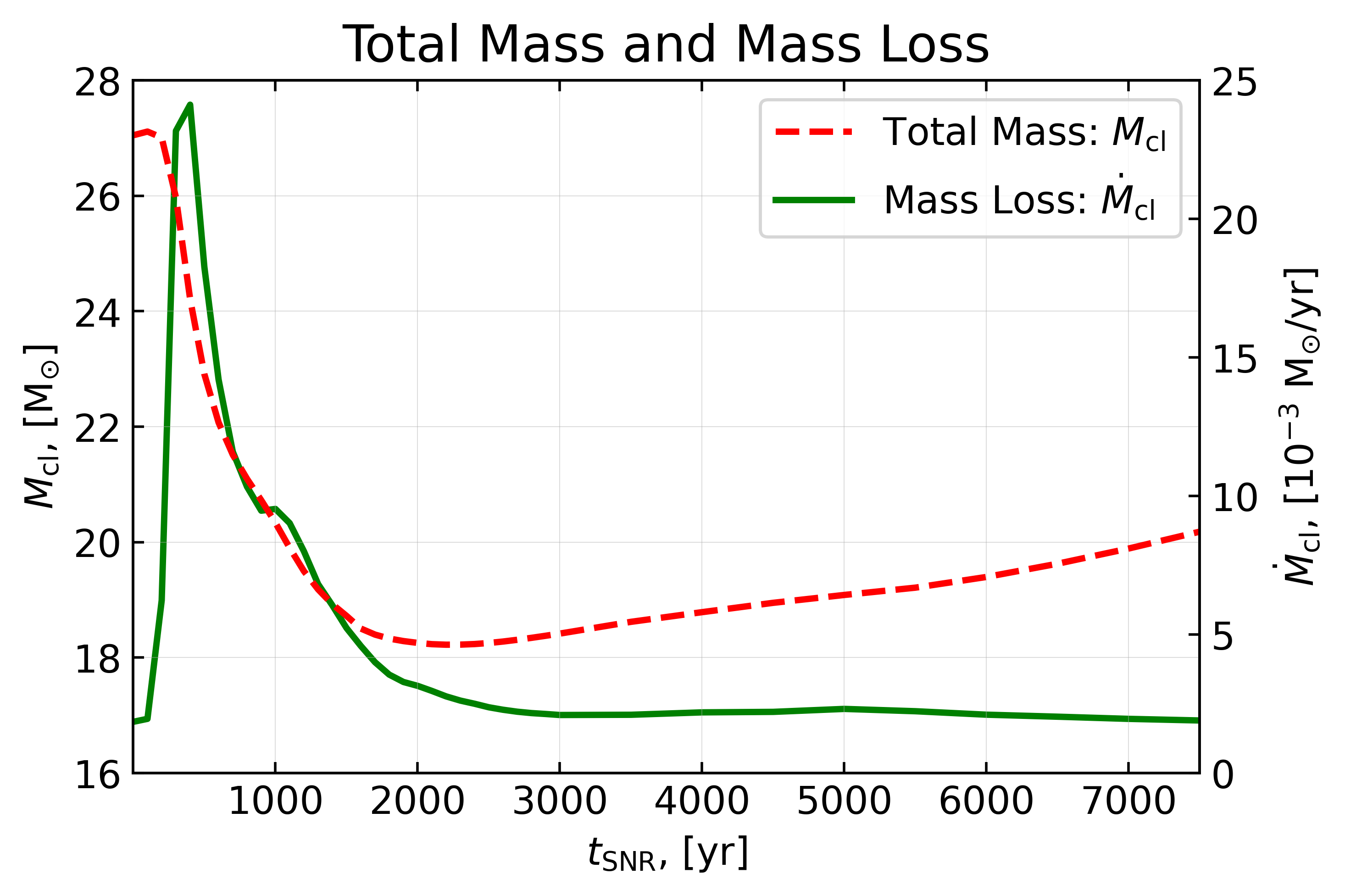}
    \vspace*{-2mm}
    \caption{Total mass of the gas in the cluster and its total mass loss rate through the cubic domain borders as functions of time $t_{\mathrm{snr}}$. The left picture stands for the centrally located SN and the right one is for the peripheral SN event.}
    \label{fig:mass}
\end{figure*}

During the first $\sim1000$ yr when the shock wave from a supernova propagates through the cluster, the ejecta replaces interstellar matter in about 50-80 per cent of the cluster core volume, depending on the position of the SN event, see Figs.~\ref{fig:mass_tr} and \ref{fig:Ejecta_pic}. A significant amount of this ejected material remains within the cluster core for $\sim3000$ years after the SN event, when 10 \Msun of material were expelled. It is interesting to note that $\approx2000$ yr after the SN event there is a total of 17-18 \Msun of diffuse gas within the cluster core, of which only $\sim0.7$-1 \Msun is the ejected material. Even though it is heavily diluted, this material still occupies about 25-30 per cent of the volume, see Fig.~\ref{fig:mass_tr}. The rate at which the cluster loses mass remains relatively stable at around $\sim2.5\times10^{-3}$\MsunYr during the considered stage of evolution. However, when the SNR shock front reaches the domain boundaries and the dense shell created by the SN starts to leave the volume, the mass loss rate increases by an order of magnitude. As a result, the total mass of gas within the cluster decreases rapidly, but then begins to grow linearly due to the effective CSG-wind mass loss, see Fig.~\ref{fig:mass}. After $t_{\rm snr}\sim4000$ yr there is almost no ejecta left inside the cluster core. Additionally, we conclude that the SN shock wave has a weak impact on the structure of neighboring heavy CSG-wind shells within a timescale of 10 kyr, compare Figs.~\ref{fig:perif} and \ref{fig:relax}.

In Fig.~\ref{fig:mass_tr} we calculated the ejecta mass fraction using passive scalar tracer $Q$, which was set to be negative for the ejecta material (see \S\ref{sec:init}): 
\begin{equation}
    M_{\rm ej}(t)=\sum_{\substack{i,j,k}}\rho_{ijk}(t)V_{\rm cell},\hspace{10pt}(i,j,k):\hspace{2pt}Q_{ijk}(t)<0.
\end{equation}
The ejecta volumetric filling factor was treated following Eq.~\ref{eq:filfac} with the above condition for $(i,j,k)$.

\subsection{X-ray diffuse emission}\label{XR}

Some of the YMSCs are bright X-ray sources (e.g. Westerlund 1, Westerlund 2). MHD simulations presented in this paper and Paper I allow modelling the thermal diffuse emission of the cluster core. Together with the analysis of the X-ray spectra of YMSCs it allows revealing the nature of their X-ray radiation, namely distinguishing between thermal and non-thermal origin of the diffuse emission. In Paper I we presented the modelling of thermal X-ray spectra of the cluster core, containing tens of massive stars, but with no SNe. Here we analysed the impact of the latest SN event in a cluster on X-ray emission spectrum. 

During the first 1000-2000 years after the SN event the cluster contains several solar masses of the ejecta. It should be taken into account that the ejecta of Type Ib/c SNe has different from the ISM chemical composition, rich in metals. As the SN ejecta was marked in our simulation, we were able to calculate the thermal spectra, considering the different chemical composition in different parts of the cluster. While for the stellar winds we used the standard table of abundances given in \cite{Asplund2009}, which is implemented in APEC model of the XSPEC, for the ejecta material we constructed the new table of abundances of elements up to Zn using the functionality of VVAPEC. We got the values of elemental abundances in SN Type Ib/c ejecta from \cite{Limongi2018}, where the detailed  modelling of stellar nucleosynthesis is presented and SN ejecta yields are found.

\begin{figure}
    \centering
    \includegraphics[width=8.25cm]{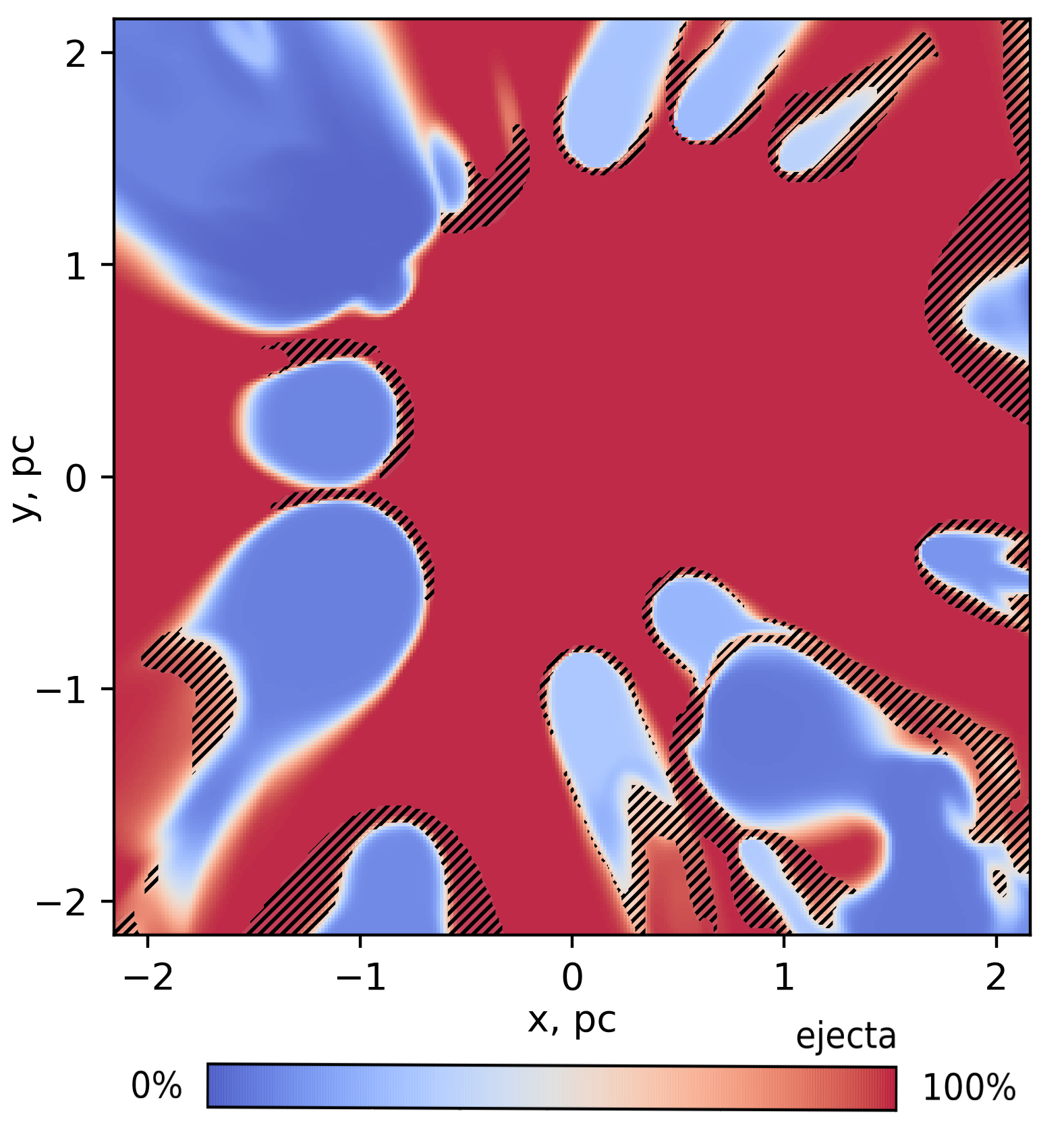}
    \vspace*{-2mm}
    \caption{The distribution of the hot ejecta of the central SN in the cluster - the material, consisting of $\gsim$ 50 \% ejecta and with $T>1$\,keV, is marked with the hatching. The colors represent the fraction of the SN ejecta in the material - pure ejecta is marked with red colour, while pure stellar wind material is marked with blue colour. Here, $t_{\rm snr}=900$\,yr.}
    \label{fig:Ejecta_pic}
\end{figure}

Even when SN shock has left the cluster core, the ejected material can still substantially enrich with the metals the hot X-ray emitting plasma in the core as it is apparent from Fig.~\ref{fig:mass_tr} and \S\ref{ej}. Most of the ejecta filling the cluster volume is cold (see lower left panels of Fig.~\ref{fig:central}, \ref{fig:perif}), but in the areas where ejecta collides with fast O-WR winds, it is heated up to a few keV, which is shown in Fig. \ref{fig:Ejecta_pic}. These areas can affect significantly the normalization and features of the X-ray diffuse thermal spectrum of the cluster. Integrating the emission (found with APEC and VVAPEC models) from each cell of the computational domain with its own $T,~\rho$ and chemical composition ('standard' or 'ejecta'), we managed to obtain the thermal spectrum from YMSC as a whole.

The MHD model we used here is single-fluid and therefore we must set an appropriate electron temperature prescription for the spectra calculation. We assumed that the hot plasma in the fast SN shock downstream has an initial $T_{\mathrm{e}} \simeq 0.1 T_{\mathrm{p}}$ as it was argued by \citet{Raymond23,vink15}. The ions heated there to $\sim$ 10 keV have not enough time to reach the thermal equilibrium with electrons in the first $\sim$ 1000 years after the SN event. Thus the temperature of electrons in the hot plasma in the SN shock downstream $T_{\mathrm{e}} \simeq 0.1 T$ (where $T$ is plasma temperature in one-fluid model) is about keV. On the other hand, the older material, heated by the stellar wind flows before the SN event, had longer time to relax to $T_{\mathrm{e}} \simeq  T_{\mathrm{p}}$ and the electron temperature there is $T_{\mathrm{e}} \simeq T/2$ which is in keV range as well.

\begin{figure}
    \centering
    \includegraphics[width=8.25cm]{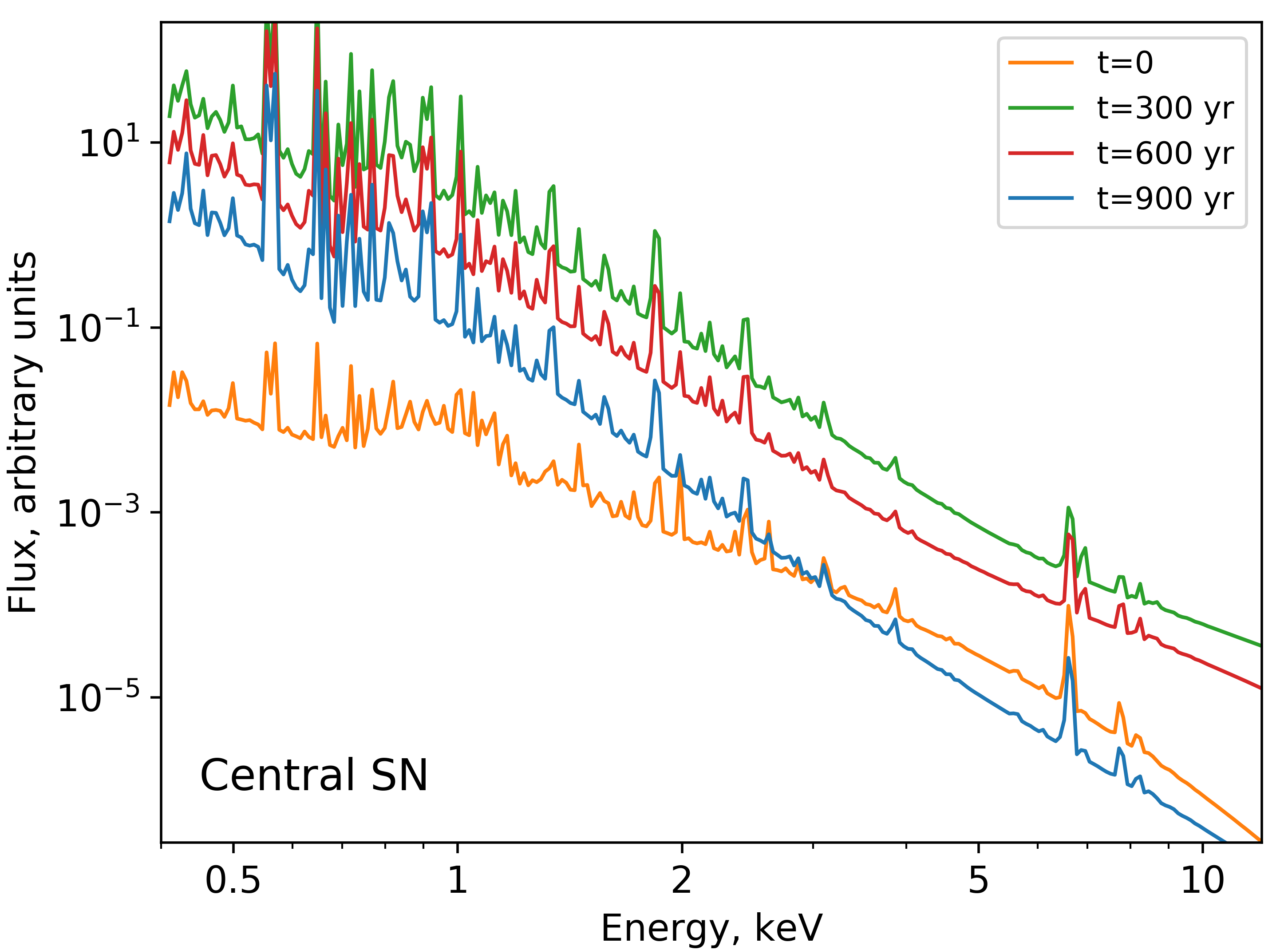}
    \vspace*{-1mm}
    \caption{Cluster thermal X-ray spectrum before the SN event (orange curve) and 300, 600 and 900 years after the event. The modelling is obtained for the central SN and the ejecta $T_e=0.1T_i$.}
    \label{fig:Xray_spec1}
\end{figure}

We present the modelled diffuse spectra in the energy range 0.5-12 keV for times up to 900 yr after the SN event in Fig. \ref{fig:Xray_spec1} where ejecta's $T_{\mathrm{e}} \simeq  0.1 T_{\mathrm{p}}$. The diffuse spectrum before the SN event (same as in Paper I) is also presented. Despite the spectra being given in arbitrary units, the relative fluxes are saved. One can see that after hundreds of years from the SN event the intensity of the X-ray emission is increased due to compression of the keV temperature plasma by SN shock. The metal-rich composition of the ejecta also leads to the increase of the X-ray flux. The decrease of the intensity with time after 300 years of the simulation is because we present the flux from a fixed domain and the compressed SN shell just moves out from the domain. The decrease of all energy profiles in Fig. \ref{fig:enrg} after $\sim 1000$ years is due to the same reason. The absence of thermal equilibrium between electrons and ions, as well as the fact that in $\sim 1000$ years most of the heated matter is swept out of the cluster, leads to the softening of the spectrum, and at high energies it can even be lower than the spectrum, calculated before the SN event.

The thermal spectra, modelled in this paper, correspond to time periods, close to the SN event and can be used for the analysis of X-ray observations of YMSCs. Such analysis for Westerlund 2 was performed in \cite{Wd2_artXC23} and it was found that in addition to thermal component, which temperature was obtained with MHD modelling, the presence of a non-thermal component is necessary to fit the {\it Chandra} and {\it ART-XC} observations. Similar discussion on the origin of the diffuse emission from other YMSCs, e.g. Westerlund 1, is needed and may be the subject of the future work.

\subsection{Nonthermal particle acceleration and radiation}

The knowledge of magnetic field structure is particularly important for modeling of very high energy particle acceleration, their confinement and the synchrotron radiation of the accelerated leptons. About $30\%$ of the cluster volume is filled with the low magnetic fields $<3$ $\mu G$ and more than a half of the volume has the field magnitude below 10 $\mu G$. The intracluster plasma is perturbed by multiple shocks and the energy-containing MHD motions of a parsec scale, which are supported by the stellar winds and supernovae.  Particle acceleration timescale in kinetic models of CR acceleration in compact clusters and superbubbles (which differ by scale sizes) can be estimated as $\tau_{\rm a}(p) \approx 9D(p)/u_{l}^2$ \citep[see e.g.][]{BK22}, where $D(p)$ is the diffusion coefficient of a CR particle of momentum $p$, and the velocity dispersion of the bulk compressible plasma motions in the system $u_{l}$ is $\gtrsim1000$\kms. The maximal energies of CR accelerated by the Fermi mechanism in the system are limited by the CR loss time $\tau_{l}(p)$ which is the shortest among the CR escape time or the synchrotron-Compton losses time (for leptons). One may get an estimation for the maximal momentum $p_{\rm max}$ of accelerated CRs from the equation $\tau_{\rm a}(p_{\rm max}) = \tau_{l}(p_{\rm max})$. Using the approximation of the electron energy loss rate with account taken of the Klein-Nishina suppression of the Compton losses on the intense optical radiation field following \citet{Moder05} one can get $c p_{\rm max} \gtrsim 10$ TeV for a cluster like Westerlund 1 or Westerlund 2. The accelerated leptons of these energies can radiate keV regime X-ray synchrotron radiation in the filaments of high magnetic field magnitudes $\gtrsim100$ $\mu G$ which fill about $5\%$ of the cluster volume and dominate the magnetic energy of the system. In \cite{Wd2_artXC23} we performed such calculation for Westerlund 2, which allowed us to explain the presence of a non-thermal component in its diffuse X-ray emission. Also, when SN shocks are passing through the close vicinity of a hot luminous star, the inverse Compton radiation of MeV electrons accelerated at the shock can contribute to the X-ray emission of the cluster. Thus, the clusters are expected to be the sources of non-thermal X-ray radiation. The hot keV plasma in the cluster also emits in X-ray energy range, which was observed in a number of clusters \citep[see e.g.][]{Mun06,Wd2_Townsley19,Kavanagh20,Wd2_artXC23} and is discussed in \S\ref{XR}.

\section{Conclusions}
\label{sec:con}

We present the 3D MHD simulations of the structure and evolution of the plasma flows, temperature and magnetic fields in the core of a young massive star cluster which is disturbed by an SN event. Both central and peripheral positions of a core-collapsed SN of energy $10^{51}$ ergs and 10 \Msun of mass ejected in a wind driven circumstellar medium are considered. The model concerns young massive clusters of a few million years age.

The SNR forward shock crushes through the YMSC core, changes its elemental abundances, heats the gas, and creates multiple bow shocks. SWs form magnetized cometary structures which penetrate into the SNR shell leaving heated up dents on its surface. This highly violent plasma state is relaxed to the quasi-stationary state over a period of $\sim$ 4000-5000 yr, while for the magnetic field structure it takes substantially longer. No significant differences in terms of densities, temperatures, and magnetic field strength are detected for the case of peripheral SN over the central one.

The structure of the cluster's perturbed magnetic fields is highly intermittent and shows a bunch of evolving elongated magnetic filaments of high magnitudes $\gtrsim$ 100 $\mu$G (up to $\sim750$ $\mu$G) and the extended regions with magnetic fields of a few 10 $\mu$G. During the propagation of the SNR shock, the range of magnetic field magnitudes $|\vect{B}|\in\left(3;30\right)$ $\mu$G becomes spatially dominant, yet the magnetic energy is mainly contributed by the fields of amplitudes $\gtrsim100$ $\mu$G with a few per cent volumetric filling factor. The presence of highly amplified magnetic fields allows the production of non-thermal X-ray synchrotron radiation by multi-TeV particles accelerated by shocks within the cluster. 

The metal-rich material ejected by core-collapse supernova reaches the maximal volumetric factor of 50-80\% (depending on the SN position) inside the cluster in about a 1000 yr after the SN event, which then decreases to $<5$\% in the next 3000 yr. Some relatively minor fraction of the SN ejecta material is heated to temperatures above a keV within the cluster core region affecting the thermal X-ray spectrum. Right after the SN event the thermal radiation (1) is increased due to the growth in thermal energy released in a cluster (2) has spectral features, connected with the mixing of a cluster material with the metal-rich SN ejecta.

\section*{Acknowledgements}
The authors wish to thank the referee for a careful reading of the paper and constructive comments. This research made use of \textmc{pluto} public MHD code developed by A. Mignone and the \textmc{pluto} team. We acknowledge the use of  data provided by NASA ADS system and SIMBAD database, operated at CDS, Strasbourg, France. Some of the modeling was performed at the JSCC RAS and the 'Tornado' subsystem of the St.~Petersburg Polytechnic University super-computing centers. 3D MHD modeling of the cluster plasma flows structure by D.V.B. and A.M.B. was supported by RSF 21-72-20020 grant. The X-ray spectra modeling was performed by M.E.K. under support from the baseline project 0040-2019-0025 at Ioffe Institute.

\section*{Data Availability}

The output data may be provided upon a reasonable request. 


\bibliographystyle{mnras}
\bibliography{mnras_template} 


\bsp	
\label{lastpage}

\end{document}